\documentclass[fp,twocolumn]{jpsj3}
\usepackage{amsmath,amssymb}
\usepackage{mathrsfs}
\def\v#1{\mib #1}

\def\lend{L_{\rm e}}
\def\sed{S_{\rm e}}
\newcommand{\aver}[1]{\left\langle {#1} \right\rangle}
\def\Jad{J_{\rm ad}}
\def\JA{J_{\rm A}}
\def\JF{J_{\rm F}}
\def\Jl{J_{\rm L}}

\def\stot{S_{\rm tot}}
\def\sztot{S^z_{\rm tot}}
\def\H{{{\cal H}}}
\def\HL{{{\cal H}_{L}}}
\def\HLo{{\cal H}_{L}^{\rm o}}
\def\HLe{{\cal H}_{L}^{\rm e}}
\def\Hsu#1{{\cal H}_{L}^{\rm #1}}

\title
{
Topological Phases of the Spin-1/2 Ferromagnetic-Antiferromagnetic Alternating Heisenberg Chain with Frustrated  Next-Nearest-Neighbour Interaction}

\author
{
Kazuo {\sc Hida},\thanks{E-mail: hida@mail.saitama-u.ac.jp}
Ken'ichi {\sc Takano},$^{1}$ and  
Hidenori {\sc Suzuki}$^{2}$
}

\inst
{Division of Material Science, Graduate School of Science and Engineering, \\ Saitama University, Saitama, Saitama, 338-8570\\ 
$^{1}$Toyota Technological Institute, 
Tenpaku-ku, Nagoya 468-851\\
$^{2}$Department of Physics, College of Humanities and Sciences, Nihon University, Setagaya-ku, Tokyo 156-8550}

\recdate
{
}

\abst
{
The spin-1/2 ferromagnetic-antiferromagnetic alternating Heisenberg chain with ferromagnetic next-nearest-neighbour (NNN) interaction is investigated. The ground state 
 is the Haldane phase for weak NNN interaction, and is the ferromagnetic phase for weak antiferromagnetic interaction.  
We find a series of topologically distinct spin-gap phases
 with various magnitudes of edge spins  for strong NNN interaction. 
The phase boundaries between these phases are determined on 
 the basis of  the DMRG calculation with additional spins that compensate the edge spins. 
 It is found that each of the exact solutions with short-range antiferromagnetic correlation on the ferromagnetic-nonmagnetic phase boundary is  representative of each spin gap phase.}

\kword
{frustration, alternating chain, Haldane phase, topological spin-gap phase, edge spin, exact solution, DMRG
}

\begin{document}
\sloppy
\maketitle
\section{Introduction}

Physics of the topological phases of matter is one of the most attractive fields of contemporary condensed matter physics. Among them, the Haldane phase of the integer-spin antiferromagnetic Heisenberg chain\cite{hal1,hal2} {is} 
 one of the simplest but nontrivial topological phases in quantum magnetism.\cite{Pollmann2010,Pollmann2012,Zang2010,Hirano2008,Chen2011} 
 As an extension of the concept of the Haldane phase, one of the authors introduced the spin-1/2 Heisenberg chain, which has two different alternating exchange interactions $\JA$ and $\JF$. The ground state of this  model interpolates those of the uniform spin-1/2 
 and spin-1 Heisenberg chains.\cite{kh} As long as  $\JA >\JF$  and $\JA$ is antiferromagnetic, this ground state is adiabatically connected to the Haldane phase of the spin-1 antiferromagnetic Heisenberg chain {without closing the bulk energy gap.} 
Therefore, this ground state belongs to the same topological phase as the Haldane state. 

\begin{figure}[h]
\centerline{\includegraphics[width=6cm]{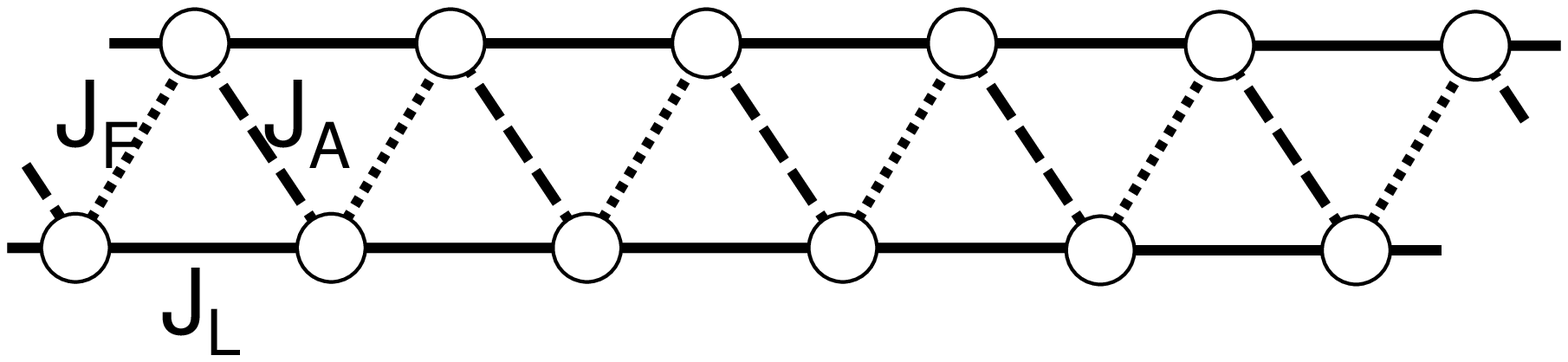}}
\caption{Structure of the frustrated ferromagnetic-antiferromagnetic alternating Heisenberg chain.}
\label{lattice}
\end{figure}

The quantum magnetism in frustrated spin systems is another rapidly developing field of condensed matter physics.\cite{diep,intfrust} Although topological phases in two-dimensional magnetic systems are often expected in  frustrated quantum magnets,\cite{huang,dep} the effect of frustration on the 
 one-dimensional topological phases has been less studied. In the present work, we investigate the effect of frustration on the ground state of  the spin-1/2 ferromagnetic-antiferromagnetic alternating Heisenberg chain. The frustration is introduced by the next-nearest-neighbour (NNN) ferromagnetic interaction $\Jl$. The lattice structure is shown in Fig. \ref{lattice}. This structure can also be regarded as a frustrated ladder\cite{hik.prb10,hako,hi} with a ferromagnetic leg interaction $\Jl$, a ferromagnetic rung interaction $\JF$, and an antiferromagnetic diagonal interaction $\JA$.  A frustrated ferromagnetic ladder compound with this structure has been synthesized recently.\cite{kato}  Although the ground state of this material is ferromagnetic, the frustration effect would be enhanced if the exchange interactions can be modified. As a result, the ground state can change into more exotic nonmagnetic phases owing to the interplay of frustration and quantum fluctuation. 
Hence, it is worthwhile to investigate the possible ground states of this model by varying the exchange interactions freely. 
\begin{figure}
\centerline{\includegraphics[width=6cm]{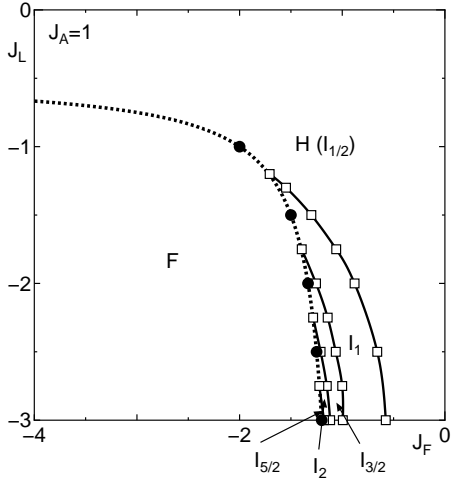}}
\caption{Ground-state phase diagram. The ferromagnetic phase, the Haldane phase, and the intermediate spin-gap phases with edge spin $\sed$ in the open chain Hamiltonian (\ref{ham_open}) are indicated by F, H, and I$_{\sed}$, respectively. The stability limit of the ferromagnetic phase (\ref{insta}) is shown by the dotted line. The open squares are data for the boundary between different spin-gap phases determined numerically in \S \ref{interm}. The filled circles correspond to the ``special points'' defined by Dmitriev {\it et al.}\cite{dmitriev,dmitriev2} The solid curves are  guides for the eye.}
\label{phase}
\end{figure}

The main results of the present study are summarized in the phase diagram of Fig. \ref{phase}. In addition to the ferromagnetic (F) phase and the Haldane (H) phase, we find  intermediate phases with various magnitudes of edge spin $\sed$  {in the region with strong ferromagnetic NNN interaction $\Jl$. The difference in the} magnitude of edge spin reflects the topological {difference} 
 of the bulk ground states.\cite{Pollmann2010,Pollmann2012,Zang2010,Hirano2008} A well-known example is the {edge spins with magnitude $1/2$ in the } Haldane phase of the spin-1 Heisenberg chain.\cite{kennedy} This state is topologically distinct from the trivial spin gap phases such as the dimer phase or the large-$D$ phase which has no edge spins.\cite{Pollmann2010,Pollmann2012,Hirano2008} The points where the exact solutions are available\cite{dmitriev,dmitriev2} are  shown also in Fig. \ref{phase}.

This paper is organized as follows. In the next section, the model Hamiltonian is presented. 
We start  with the classical analysis {in \S 3}, and then determine the ground-state phase diagram by the numerical diagonalization and DMRG technique in \S 4. The relation to the exact solution on the ferromagnetic-nonmagnetic phase boundary is also discussed. The last  section is devoted to a summary and discussion.
\section{Hamiltonian}
\label{section:ham}
We aim to clarify the nature of the ground state of the spin-1/2 Heisenberg chain described by the Hamiltonian
\begin{align}
{\HL} &=\sum_{l=1}^{L} \left(\JF\v{S}_{2l-1}\v{S}_{2l}+\JA\v{S}_{2l}\v{S}_{2l+1}\right)+\sum_{i=1}^{2L}\Jl\v{S}_{i}\v{S}_{i+2}, \label{hama}
\end{align}
where $\v{S}_{l}$  is the spin-1/2 operator on the $l$-th site. 
In the present work, we concentrate on the case where $\JF , \Jl < 0$ (ferromagnetic) and $\JA > 0$ (antiferromagnetic).
 If we assume the periodic boundary condition $\v{S}_{l+2L}=\v{S}_l$,  the total number of spins is $2L$. In the following sections, however, we also consider several different open chains with additional spins suitable for elucidating the nature of different phases and phase transitions between them.
\section{Classical Ground-State Phases}
Before analyzing the quantum ground states of Hamiltonian (\ref{hama}), we examine the ground states of its classical version for comparison. 
The classical Hamiltonian 
$\Hsu{cl}$ is given by replacing the spin 1/2 operator $\v{S}_{i}$ by a classical vector with length $S$ in ${\HL}$. 

\subsection{Ferromagnetic phase}

We rewrite 
{$\Hsu{cl}$} in the following complete-square form 
\begin{align}
{\Hsu{cl}} = \frac{1}{2} \sum_{l=1}^{L} &\bigg[
A\left\{(\v{S}_{2l-1}-\lambda\v{S}_{2l}-(1-\lambda)\v{S}_{2l+1})^2\right.\nonumber\\
&\left.+(\v{S}_{2l+2}-\lambda\v{S}_{2l+1}-(1-\lambda)\v{S}_{2l})^2\right\}\nonumber\\
&+B(\v{S}_{2l}-\v{S}_{2l+1})^2\bigg]+E_{\rm F} , 
\end{align}
where $\lambda=\JF/(\JF+2\Jl)$, 
$A=-(\JF+2\Jl)/2$, $B=-2\Jl\JF/(\JF+2\Jl)-\JA$, 
and 
\begin{align}
E_{\rm F} =LS^2 ({2\Jl}+{\JF}+{\JA}).
\end{align}

As long as  $A>0$ and $B>0$, 
the ground state is the ferromagnetic state 
where $\v{S}_{i}$ is independent of $i$, 
and the ground-state energy is $E_{\rm F}$. 
$A$ is always positive for $\JF <0$ and $\Jl <0$.
Hence the condition $B>0$ written as
\begin{align}
-\frac{1}{2\Jl}\leq \frac{1}{\JF}+\frac{1}{\JA} 
\label{ferro_class}
\end{align}
gives the ferromagnetic phase.
\subsection{Antiferromagnetic phase}
We rewrite $\Hsu{cl}$ in the following complete-square form 
\begin{align}
\Hsu{cl} = \frac{1}{2} \sum_{l=1}^{L} &\bigg[
A\left\{(\v{S}_{2l-1}+\lambda\v{S}_{2l-2}+(1-\lambda)\v{S}_{2l})^2\right.\nonumber\\
&\left.+(\v{S}_{2l}+\lambda\v{S}_{2l+1}+(1-\lambda)\v{S}_{2l-1})^2\right\}\nonumber\\
&+B(\v{S}_{2l-2}+\v{S}_{2l-1})^2 \bigg]+E_{\rm AF} ,
\end{align}
where $\lambda=2\Jl/\JF$, 
$A=-\JF^2/[2(2\Jl-\JF)]$, 
$B=\JA+2\Jl\JF/(2\Jl-\JF)$, 
and 
\begin{align}
E_{\rm AF}=LS^2(2\Jl-\JF-\JA).
\end{align}

For $A \geq 0$ and $B \geq 0$, 
 the ground state is the antiferromagnetic (AF) state where $\v{S}_{l+1}=-\v{S}_{l}$ for all $l$, and the ground state energy is $E_{\rm AF}$. 
These conditions reduce to 
\begin{align}
\frac{1}{2\Jl}\geq \frac{1}{\JF}+\frac{1}{\JA}
\label{af_class}
\end{align}
for the antiferromagnetic (AF) phase.
\subsection{Double-period antiferromagnetic phase}
We rewrite $\Hsu{cl}$ in the following complete-square form  
\begin{align}
\Hsu{cl} = \frac{1}{2} \sum_{l=1}^{L} &\bigg[
A\left\{(\v{S}_{2l-2}+\lambda\v{S}_{2l-1}+(1-\lambda)\v{S}_{2l})^2\right.\nonumber\\
&\left.+(\v{S}_{2l-1}+\lambda\v{S}_{2l-2}+(1-\lambda)\v{S}_{2l-3})^2
\right\}\nonumber\\
&+B(\v{S}_{2l-2}+\v{S}_{2l-1})^2
\bigg]+E_{\rm DAF}
\end{align}
where $\lambda=\JF/(2\Jl)$, 
$A=2\Jl^2/(2\Jl-\JF)$, 
$B=\JA-2\Jl\JF/(2\Jl-\JF)$, 
and the ground-state energy $E_{\rm DAF}$ is given as 
\begin{align}
E_{\rm DAF}=-LS^2(2\Jl-\JF+\JA).
\end{align}

For $A \geq 0$ and $B \geq 0$, 
 the ground state is the antiferromagnetic state with doubled periodicity (DAF state) where $\v{S}_{2l}=\v{S}_{2l-1}=-\v{S}_{2l-2}=-\v{S}_{2l-3}$ 
 for all $l$, and the ground-state energy is $E_{\rm DAF}$. 
These conditions reduce to
\begin{align}
-\frac{1}{2\Jl}\geq -\frac{1}{\JF}+\frac{1}{\JA}
\label{period4_class}
\end{align}
for the double-period antiferromagnetic (DAF) phase.

\subsection{Spiral phase}

We rewrite  
$\Hsu{cl}$ in the following complete-square form  
\begin{align}
\Hsu{cl} = \frac{1}{2} &\sum_{l=1}^{L} 
\left\{(\alpha\v{S}_{2l-1} + \beta\v{S}_{2l} + \gamma\v{S}_{2l+1})^2\right.\nonumber\\
&\left.+(\gamma\v{S}_{2l} + \beta\v{S}_{2l+1} + \alpha\v{S}_{2l+2})^2\right\} +E_{\rm S} , 
\label{spiral_ham}
\end{align}
with 
$\alpha = \sqrt{\JA \Jl/\JF}$, 
$\beta = (1/2)\sqrt{\JA \JF/\Jl}$, 
$\gamma = -\sqrt{\Jl \JF/\JA}$, 
and 
\begin{align}
E_{\rm S}&=-LS^2 \left( \frac{\Jl \JA}{\JF} 
+ \frac{\JF \JA}{4\Jl} + \frac{\Jl \JF}{\JA} \right) . 
\end{align}

A ground-state spin configuration is such that 
 all the squares in eq.~(\ref{spiral_ham}) are zero  and 
 the ground state energy is $E_{\rm S}$. 
In the ground state, both $\v{S}_{2l-1}$ and $\v{S}_{2l+2}$ are on the plane spanned by $\v{S}_{2l}$ and $\v{S}_{2l+1}$, and then
 all the spins are in the same plane. 
Setting the plane as the $xy$-plane, the spin configuration is expressed 
 as 
\begin{align}
\v{S}_{n}&=(S\cos\theta_n, S\sin\theta_n, 0) , 
\end{align}
where 
\begin{align}
\theta_{2l} &= \theta_0 + l\theta , \nonumber\\
\theta_{2l+1} &= \theta_0 + (l+1)\theta + \phi 
\end{align}
with 
\begin{align}
\phi &= \cos^{-1} \frac{\gamma^2 -\alpha^2 -\beta^2}{2\alpha\beta} , 
\nonumber\\
\theta &= -\phi + \cos^{-1} \frac{\alpha^2 -\beta^2 -\gamma^2}{2\beta\gamma} ,
\label{spiral_sol}
\end{align}
and arbitrary $\theta_0$. 
This spin configuration shows that the system is in the spiral (S) phase with period $\theta$. 
The ground state is a commensurate spiral state if $\theta/\pi$ is a rational number, and otherwise, an incommensurate one. 
For example, in the case of $\Jl = -\JA/2$ and $\JF = -\JA$, we have a commensurate spiral state with $\theta = \phi = \pi/3$ and 
$E_S = -(3/2)\JA {S^2L}$. 

The spiral spin configuration exists if and only if $|\cos \phi| \le 1$ and $|\cos(\theta + \phi)| \le 1$ for eq. (\ref{spiral_sol}). 
These conditions restrict the values of the exchange parameters and determine the phase boundaries between the spiral phase and the other phases. 
Rewriting the conditions using the exchange parameters, we find that the spiral phase occupies 
 all the area outside 
 the F, AF, and DAF phases respectively given by eqs.~(\ref{ferro_class}), (\ref{af_class}), and (\ref{period4_class}). 
The spiral spin configuration reduces to the F, AF, and DAF configurations on the boundaries, as expected.

\begin{figure}
\centerline{\includegraphics[width=6cm]{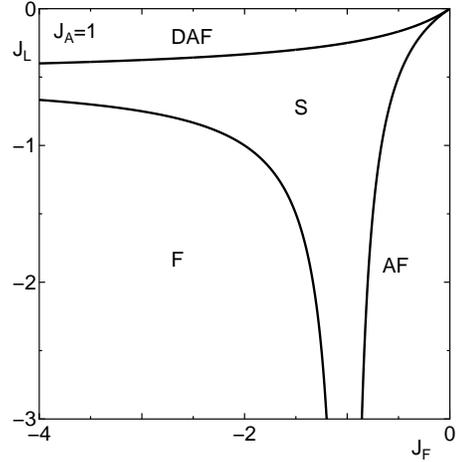}}
\caption{Classical ground-state phase diagram. 
It includes the ferromagnetic (F), antiferromagnetic (AF), double-period antiferromagnetic (DAF), and spiral (S) phases. 
}
\label{phase_class}
\end{figure}

The classical ground-state phase diagram is summarized in Fig. \ref{phase_class}.
A comparison of this classical phase {diagram} to the quantum one in Fig. \ref{phase} reveals that the ferromagnetic phase 
 remains in 
 the same area. 
The spiral phase together with the AF and DAF phases seems to turn into the  Haldane phase with spin gap and short-range order owing to quantum fluctuation. 
The intermediate phases between the ferromagnetic and Haldane phases are proper to the quantum system, and will be examined in detail in the next section. 
\section{Ground-State Phase Diagram}
\subsection{Strong coupling limit :  $|\Jl|, \JA \ll |\JF|$}\label{strong}
In the strong $\JF$ limit, $\v{S}_{2l-1}$ and $\v{S}_{2l}$ form a triplet pair that is described by the spin-1 operator $\hat{\v{S}}_l=\v{S}_{2l-1}+\v{S}_{2l}$. Then the original Hamiltonian 
(\ref{hama}) 
is mapped onto an effective spin-1 Heisenberg chain, 
\begin{align}
{\cal H}_{L}^\mathrm{eff} &=\sum_{l=1}^{L} J_{\rm eff}\hat{\v{S}}_{l}\hat{\v{S}}_{l+1}, 
\label{hameff}
\end{align}
with an effective exchange parameter $J_{\rm eff} = (2\Jl + \JA)/4$. 
The ground state is 
a ferromagnetic state for $\Jl < -\JA/2$, 
whereas it is a 
Haldane state for $\Jl > -\JA/2$.

{
{As long as the spins $\v{S}_{2l-1}$ and $\v{S}_{2l}$ form a composite spin $\hat{\v{S}}_l$ with magnitude 1, 
 }
the Haldane state has an unambiguously
 topological nature.\cite{Pollmann2010,Pollmann2012,Zang2010,Hirano2008} Reflecting this fact, the edge spins with magnitude 1/2 
 appear at the open edges in the Haldane phase. 
However, actual {spins} $\v{S}_{2l-1}$ and $\v{S}_{2l}$ {in Hamiltonian (\ref{hama})} are {independent degrees of freedom, 
 although they interact with each other.} 
If we cut 
{the chain }
between $\v{S}_{2l-1}$ and $\v{S}_{2l}$, no edge spins appear at the {boundary} 
 in the Haldane phase. Therefore, {we can also regard this phase 
 as a trivial spin-gap phase. }
Thus, the definitions of the terms ``topological'' and ``trivial'' are rather arbitrary in the present model. 
To fix 
 the terminology, we choose the pairs of spins on the $\JF$ bonds as  building blocks of the bulk of our chain and call the phase that adiabatically continues to the Haldane phase of the spin-1 chain in the limit $\JF \rightarrow -\infty$ with $\Jl < -\JA/2$  ``topological''. 
}
\subsection{Ferromagnetic-nonmagnetic phase boundary}

The stability limit of the ferromagnetic phase can be determined from the requirement that the single magnon excitation energy vanishes\cite{hi} as
\begin{align}
   \JF&=\JF^{\rm s}\equiv-\frac{2\JA\Jl}{2\Jl+\JA}.
\label{insta} 
\end{align}
This is plotted as the dotted line in Fig. \ref{phase}. This 
is the same as the classical phase boundary given by eq. (\ref{ferro_class}). It is also numerically confirmed that no partial ferromagnetic phases\cite{sachdev,ir,ym,khferri,shimo1,
hts,htsdec,khdlt} appear between the ferromagnetic 
and nonmagnetic phases. Hence, this transition is the first-order transition accompanied by the discontinuous change of the total magnetization.

\subsection{Intermediate spin-gap phases}\label{interm}
\begin{figure}
\centerline{\includegraphics[width=6cm]{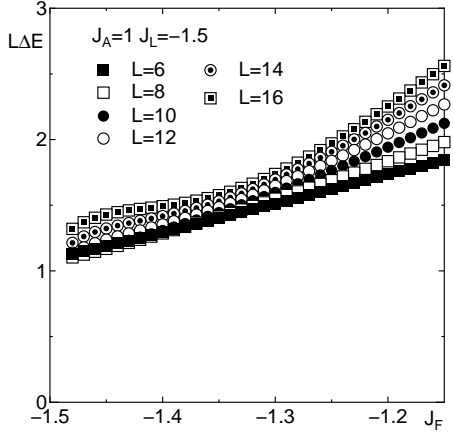}}
\caption{$\JF$-dependence of the scaled gap $L\Delta E$ with the periodic boundary condition for $\Jl=-1.5$.
}
\label{exo_peri}
\end{figure}

 {In the following numerical calculation, we fix the energy unit by setting $\JA=1$ without loss of generality}. Figure \ref{exo_peri} shows the $\JF$-dependence of the scaled lowest singlet-triplet energy gap $L \Delta E$ for $\Jl=-1.5$ and $\JA=1$ calculated by the numerical diagonalization method with the periodic boundary condition.  
The size dependence of the scaled gap $L\Delta E$ is weak around $\JF \sim -1.3$ and $L\Delta E$ increases with $L$ on  both sides of this point. This suggests that a phase transition between two different spin-gap phases takes place around this point.  Namely,  an intermediate spin-gap phase, which is different from the Haldane phase, exists between the Haldane and ferromagnetic phases.  

To gain more insight into the properties of this intermediate phase and to determine the phase boundary between the intermediate  and  Haldane phases more precisely, we employ the DMRG method {for} 
 the open chain Hamiltonian,
 \begin{align}
\HLo &=\sum_{l=1}^{L} \JF\v{S}_{2l-1}\v{S}_{2l}+\sum_{l=1}^{L-1} \JA\v{S}_{2l}\v{S}_{2l+1}+\!\sum_{i=1}^{2L-2}\!\Jl\v{S}_{i}\v{S}_{i+2},
\label{ham_open}
\end{align}
where the total number of spins is $2L$. 
This Hamiltonian naturally continues to the open spin-1 Heisenberg chain 
 in the limit of $\JF \rightarrow -\infty$. Therefore, the quasi-degeneracy of the ground state due to the edge spins with magnitude 1/2 is expected in the Haldane phase. 
In the present model, however, the lowest magnetic excitation gap is extremely small  under this boundary condition 
 in the whole interval of $\JF^{\rm s}\equiv -1.5 \leq \JF \leq 0$ for $\Jl=-1.5$. 
 This suggests that the ground states are quasi-degenerate even in the intermediate phase, although the precise analysis of the size dependence of the scaled gap is difficult because of 
the smallness of the energy gap. This is in contrast to the conventional Gaussian transition between two spin-gap phases such as the Haldane-dimer transition in dimerized spin-1 Heisenberg chains.\cite{ah,chs} 

\begin{figure}
\centerline{\includegraphics[width=6cm]{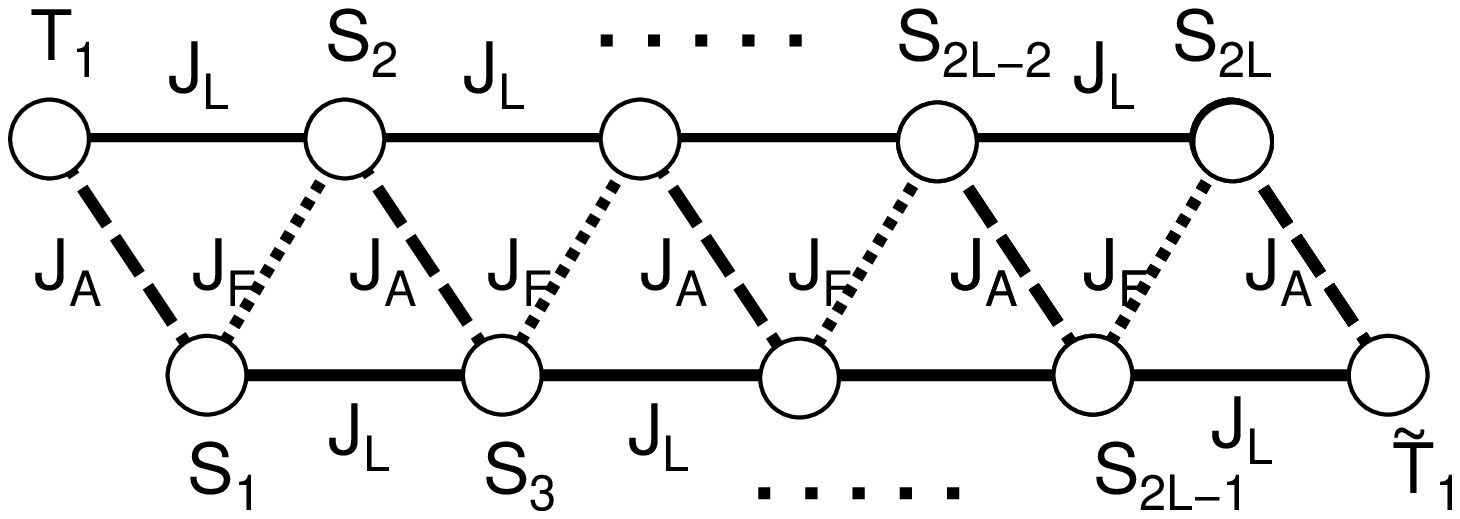}}
\caption{Structure of frustrated ferromagnetic-antiferromagnetic alternating chain described by Hamiltonian (\ref{ham_endspin}) with  spins $\v{T}_1$ and $\tilde{\v{T}}_1$ added on both ends. The total number of spins is $2L+2$.}
\label{lattice_endspin}
\end{figure}
\begin{figure}
\centerline{\includegraphics[width=6cm]{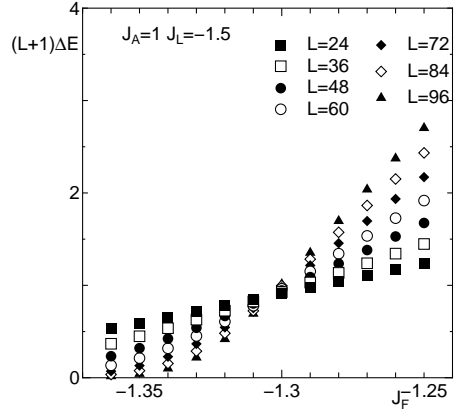}}
\caption{$\JF$-dependence of the scaled gap with $\Jl=-1.5$ 
 for the lattice geometry in Fig. \ref{lattice_endspin}.}
\label{gap_endspin}
\end{figure}

To resolve the quasi-degeneracy of the ground state and the low-lying excited states in the Haldane phase, we add two spins, $\v{T}_1$ and $\tilde{\v{T}}_{1}$, on the two ends of the chain as
\begin{align}
\HLe &= 
\HLo+
\JA(\v{T}_{1}\v{S}_{1}+\v{S}_{2L}\tilde{\v{T}}_{1})\nonumber\\
&+\Jl(\v{T}_{1}\v{S}_{2}+\v{S}_{2L-1}\tilde{\v{T}}_{1}).\label{ham_endspin}
\end{align}
This geometry is shown in Fig. \ref{lattice_endspin}. {It should be noted that the additional spins  ${\v{T}}_{1}$ and $\tilde{\v{T}}_{1}$ are merely added to compensate the edge spins in the Haldane phase to elucidate the nature of the original open chain $\HLo$ 
which tends to the spin-1 open chain in the limit of $\JF \rightarrow -\infty$. 
} In this case, the total number of spins is $2(L+1)$. The behavior of the scaled gap $(L+1) \Delta E$ is shown in Fig. \ref{gap_endspin}. It increases with the system size in the Haldane phase, as expected, whereas it still decreases with the system size in the intermediate phase. 
 This implies that the edge spins 
 in the intermediate phase are not fully compensated by the added spins $\v{T}_1$ and $\tilde{\v{T}}_1$ in  Hamiltonian (\ref{ham_endspin}). Actually, for this Hamiltonian, the lowest energy state with the {$z$-component of the total spin $\sztot=1$} and the ground state with {$\sztot=0$} are quasi-degenerate but the lowest energy state with {$\sztot=2$} is separated by a gap of $O(1)$ in the intermediate phase. Namely, the ground state is quasi-degenerate with the lowest energy state with {total spin $\stot=1$} as the Kennedy triplet in the open $S=1$ antiferromagnetic Heisenberg chain.\cite{kennedy} We also find that the ground state of the chains with no additional spins, Hamiltonian (\ref{ham_open}), is quasi-degenerate with the lowest energy state with {$\sztot=2$} in the intermediate phase. 

\begin{figure}
\centerline{\includegraphics[width=6cm]{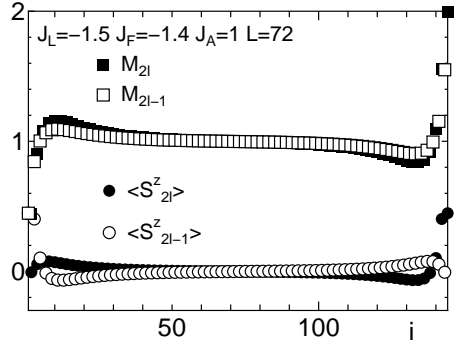}}
\caption{Magnetization profile $\aver{S^z_{i}}$ and accumulated magnetization $M_i$ for  $\Jl=-1.5$ {and} $\JF=-1.4$.  Filled and open symbols represent the values for $i=2l$ and $2l-1$, respectively.}
\label{szav}
\end{figure}

To reveal the spin structure of these states, we plot the local expectation values {$\aver{S^z_i}$} 
 in the lowest energy state of Hamiltonian (\ref{ham_open}) with {$\stot=\sztot=2$} in Fig. \ref{szav} for  $\Jl=-1.5$ {and} $\JF=-1.4$. 
 The accumulated magnetization $M_i\equiv\sum_{k=1}^{i}\aver{S^z_{k}}$ is also plotted. This plot clearly shows that  edge spins with magnitude 1 appear on the both ends of the chain. At the left (right) end, the magnetization is mostly localized on the odd-th (even-th) sites. 
 This observation also suggests the possible emergence of the phases with larger edge spins  for larger $|\Jl|$ where the odd-th (even-th) spins are strongly correlated with the leftmost (rightmost) spin ferromagnetically, owing to the next-nearest-neighbour interaction $\Jl$. 

\begin{figure}
\centerline{\includegraphics[width=9cm]{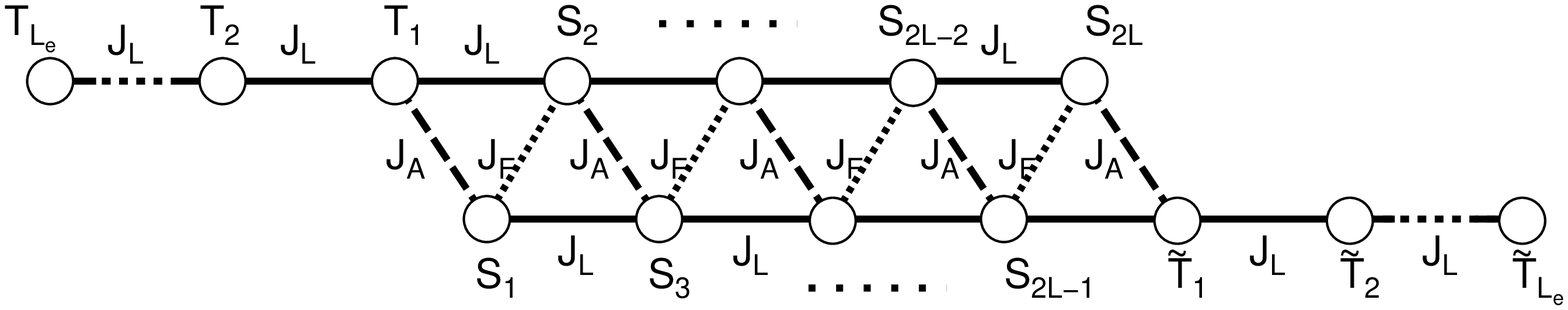}}
\caption{Structure of frustrated ferromagnetic-antiferromagnetic alternating chain described by  Hamiltonian (\ref{ham_endspin_gen}) with  spins $\v{T}_l$ and $\tilde{\v{T}}_l\ (l=1,...,\lend)$ added on both ends.  The total number of spins is $2L+2\lend$.}
\label{lattice_endspin_gen}
\end{figure}

Motivated by this speculation, we examine the Hamiltonian 
 \begin{align}
{\H}_{L \lend} &= \HLo 
+\JA\v{T}_1\v{S}_1+\Jl\v{T}_1\v{S}_2+\sum_{l=2}^{\lend}\Jl\v{T}_{l-1}\v{T}_{l}\nonumber\\
&+\JA\tilde{\v{T}}_{1}\v{S}_{2L}+\Jl\tilde{\v{T}}_1\v{S}_{2L-1}+\sum_{l=2}^{\lend}\Jl\tilde{\v{T}}_{l-1}\tilde{\v{T}}_{l}
\label{ham_endspin_gen}
\end{align}
with additional $2\lend$ ferromagnetically coupled spins $\v{T}_{l}$ and $ \tilde{\v{T}}_{l}$ to  both ends of the open chain Hamiltonian (\ref{ham_open}).  The lattice structure is shown in Fig. \ref{lattice_endspin_gen}. The total number of spins is $2L+2\lend$. 
When a ground state of $\HLo$ has edge spins with magnitude $\sed\equiv \lend/2$, they are fully compensated by the added spins in ${\H}_{L \lend}$. 
Hence, 
 each intermediate spin-gap phase of $\HLo$ is identified by the number of spins $\lend$ required to compensate edge spin.

\begin{figure}
\centerline{\includegraphics[width=6cm]{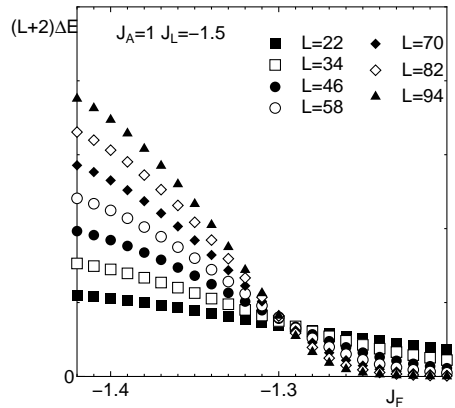}}
\caption{$\JF$-dependence of scaled gap with $\Jl=-1.5$ 
 for the lattice geometry in Fig. \ref{lattice_endspin_gen} with $\lend=2$.}
\label{gap_l30}
\end{figure}

In what follows, we call the intermediate phase that has the edge spin with magnitude $\sed$ in Hamiltonian (\ref{ham_open}), the I$_{\sed}$ phase. The I$_{1/2}$ phase is the Haldane phase. The $\JF$-dependence of the scaled energy gap is shown in Fig. \ref{gap_l30} 
for $\Jl=-1.5$  
 and $\lend=2$. In this case, the scaled gap increases with the system size in the intermediate phase, while it  decreases with the system size in the Haldane phase. This implies that the edge spins in the intermediate phase of Hamiltonian (\ref{ham_open}) are 
 fully compensated by the added spins, as expected. This confirms that the magnitude $\sed$ of the edge spin is unity in this phase. 

\begin{figure}
\centerline{\includegraphics[width=6cm]{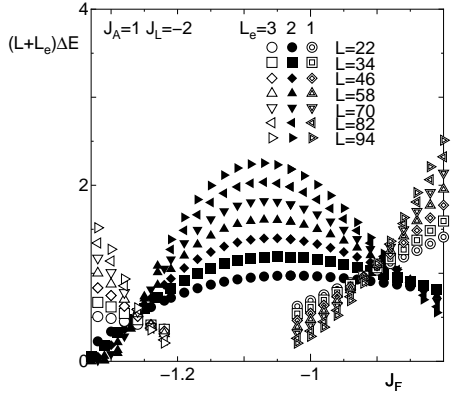}}
\caption{$\JF$-dependence of scaled gap with $\Jl=-2$ 
 for the lattice geometry in Fig. \ref{lattice_endspin_gen}. $\lend=1,2$, and 3.}
\label{gap_l40}
\end{figure}

\begin{figure}
\centerline{\includegraphics[width=6cm]{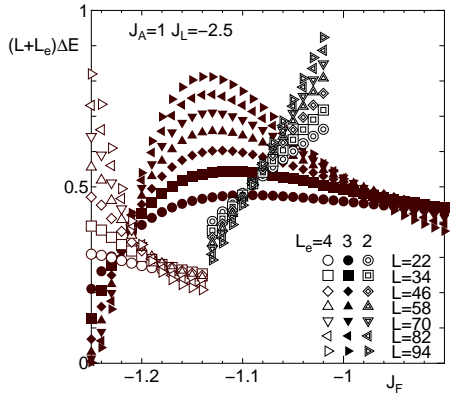}}
\caption{$\JF$-dependence of  scaled gap for $\Jl=-2.5$ 
 with the lattice geometry in Fig. \ref{lattice_endspin_gen}. $\lend=2, 3$, and 4.}
\label{gap_l50}
\end{figure}

The $\JF$-dependence of the scaled energy gap is shown in Fig. \ref{gap_l40} for $\Jl=-2$, 
 and $\lend=1, 2$, and 3. The same plot is shown in Fig. \ref{gap_l50} for $\Jl=-2.5$ and $\lend=2, 3$, and 4. The scaled energy gap increases with $L$ if the edge spins are compensated by the added spins. For $\Jl=-2.0$
, we find three phases with edge spins $\sed=\lend/2=1/2, 1$, and 3/2, whereas for $\Jl=-2.5$
, we find four phases with $\sed=\lend/2=1/2, 1, 3/2$, and 2, although the data for $\lend=1$ are not shown.  

\begin{figure}
\centerline{\includegraphics[width=6cm]{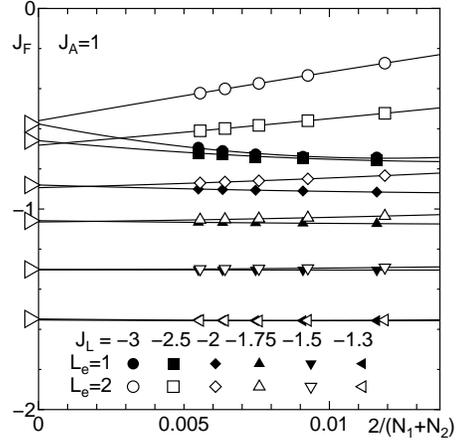}}
\caption{Extrapolation procedure for the critical values of $\JF$ for  $\Jl=-1.5$ 
 determined from the intersection point of the scaled gaps for $N=N_1$ and $N=N_2$ where $N\equiv 2L+2\lend$ is the total number of spins. The filled symbols and open symbols are the data for $\lend=1$ and $2$, respectively. The right-directed triangles on the vertical axis are the extrapolated values.}
\label{extra}
\end{figure}
\begin{figure}
\centerline{\includegraphics[width=6.5cm]{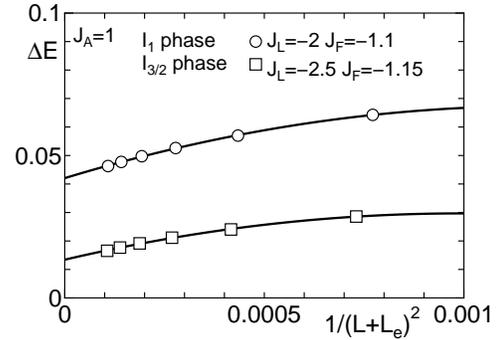}}
\caption{Size dependence of the energy gap for $\Jl=-2, \JF=-1.1$ (I$_{1}$ phase) with $\lend=2$ and  $\Jl=-2.5, \JF=-1.15$ (I$_{3/2}$ phase) with $\lend=3$. }
\label{gaps}
\end{figure}
Following the scheme of the phenomenological renormalization group,\cite{prg} we determine the finite-size phase boundary from the intersection point of the scaled gap for Hamiltonian (\ref{ham_endspin_gen}). To determine the I$_{\sed}$-I$_{\sed+1/2}$ phase boundary, we use the energy gap for ${\HL}_{\lend}$ and ${\HL}_{\lend+1}$ with $\lend=2\sed$.  After extrapolation to the thermodynamic limit $L \rightarrow \infty$, the two results are found to coincide within $2 \times 10^{-2}
$, even in the worst case, for the data shown in Fig. \ref{phase}. The extrapolation procedure is shown in Fig. \ref{extra} for the Haldane-I$_{1}$ phase boundary with $\Jl=-1.5$
. In Fig. \ref{phase}, we plotted the I$_{\sed}$-I$_{\sed+1/2}$ boundary determined from ${\HL}_{2\sed}$, because its size dependence is weaker than that determined from ${\HL}_{2\sed+1}$ empirically. 

{The size dependence  of the bulk energy gap in the middle of I$_{1}$ and I$_{3/2}$ phases is plotted against the system size in Fig. \ref{gaps}. It is seen that the magnitude of the energy gap is small but finite in these phases.}

{\subsection{Valence bond structure of the intermediate phases}}

\begin{figure}
\centerline{\includegraphics[width=8cm]{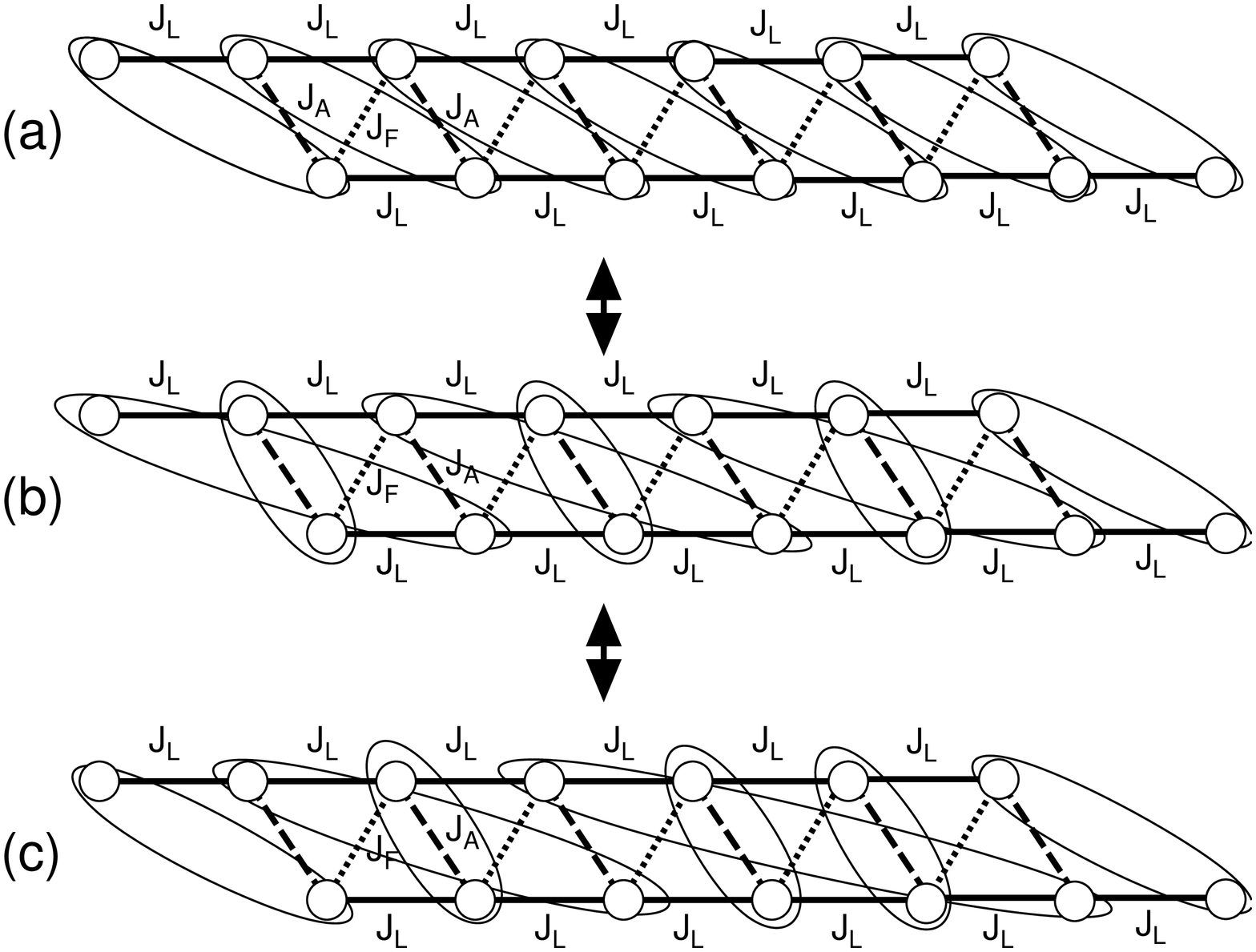}}
\caption{Structure of the I$_1$ phase with $\lend=2$. Ovals correspond to the valence bonds.}
\label{lattice_lrvbs}
\end{figure}

Since each I$_{\sed}$ phase is gapped, its valence bond structure is short-ranged. Nevertheless, the actual valence bond configuration is 
 rather enigmatic. For example, the simplest possible structure of  the I$_1$ phase speculated from the magnitude of the end spins would be the one shown in Fig. \ref{lattice_lrvbs}(a). To examine the possibility of this structure, we have measured the spin correlation functions $\aver{\v{S}_{2l+2j+1}\v{S}_{2l}}$ with $-4 \leq j \leq 3$ using DMRG near the center of the chain.

\begin{figure}
\centerline{\includegraphics[width=6cm]{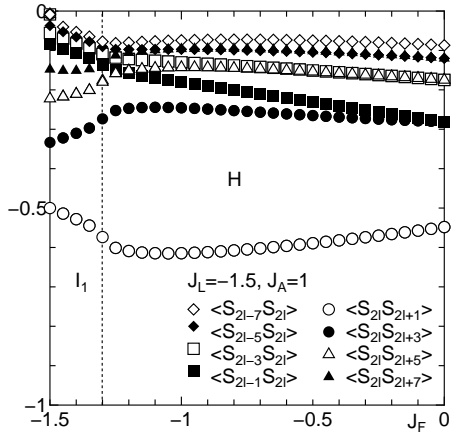}}
\caption{Short-range spin-spin correlation functions for $\Jl=-1.5$ 
 calculated by the DMRG method. The system size dependence is negligible except at the H-I$_1$ phase transition point (dotted line) where the values extrapolated to $L \rightarrow \infty$ are plotted.}
\label{corel}
\end{figure}

{Contrary to the above intuition,} we find that $\aver{\v{S}_{2l}\v{S}_{2l+1}}$ has  the largest amplitude not only in the Haldane phase but also in the I$_{1}$ phase, as shown in Fig. \ref{corel} for $\Jl=-1.5$ and $\JA=1$. Nevertheless, it decreases with $\JF$, and the correlation $\aver{\v{S}_{2l}\v{S}_{2l+3}}$ increases with $\JF$. Furthermore, a cusplike behavior is observed in all short-range correlation functions at the phase boundary. Thus, we speculate that the dimer configuration of Fig. \ref{lattice_lrvbs}(a) resonates with  other longer range  valence bond structures 
to lower the energy of the I$_{1}$ phase. Two examples of such valence bond 
 configurations are shown in Figs. \ref{lattice_lrvbs}(b) and  \ref{lattice_lrvbs}(c). 
They are compatible with the spin gap for Hamiltonian (\ref{ham_endspin_gen}) with $\lend=2$. 
\begin{figure}
\centerline{\includegraphics[width=7cm]{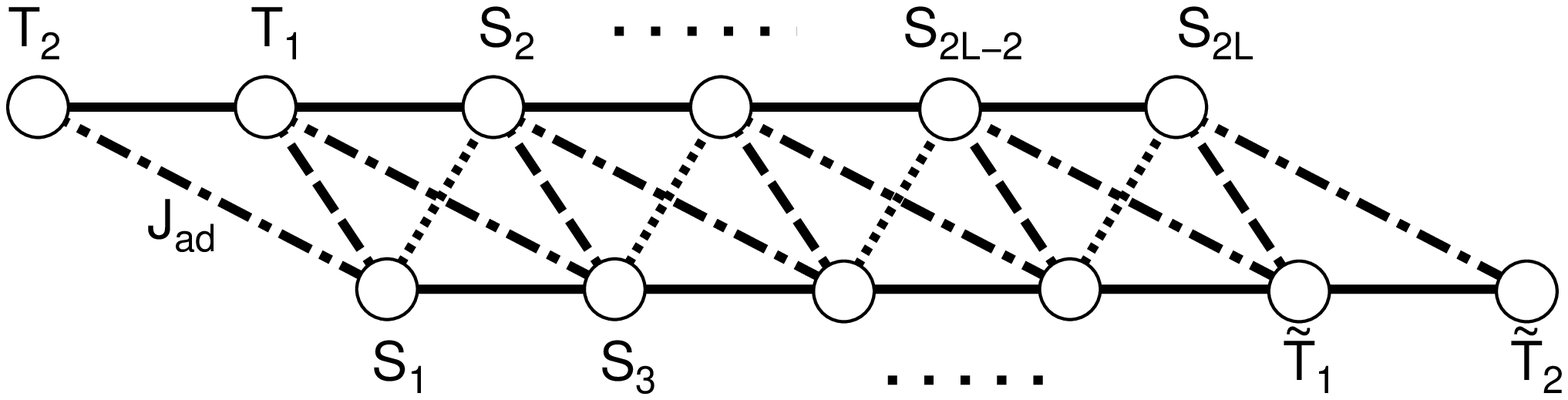}}
\caption{Structure of the frustrated ferromagnetic-antiferromagnetic alternating chain described by Hamiltonian (\ref{ham_endspin_ad}) with $\Jad$.  The total number of spins is $2L+4$.}
\label{lattice_endspin2add}
\end{figure}
\begin{figure}
\centerline{\includegraphics[width=6cm]{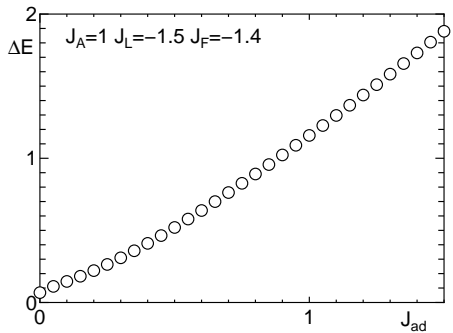}}
\caption{$\Jad$-dependence of the energy gap of  Hamiltonian (\ref{ham_endspin_ad}) with $\JA=1$, $\Jl=-1.5$, and $\JF=-1.4$ calculated by the DMRG method. The gap extrapolated to the thermodynamic limit $L \rightarrow \infty$ is shown.}
\label{gapad}
\end{figure}

To confirm this picture, we consider the model with  additional antiferromagnetic bonds $\Jad$ to {${\H}_{L\lend}$ with $\lend=2$}, as shown in Fig. \ref{lattice_endspin2add}. 
The Hamiltonian is given by 
 \begin{align}
{\Hsu{ad}} &={\H}_{L 2}+\Jad\sum_{l=1}^{L-2}{\v{S}}_{2l}{\v{S}}_{2l+3}\nonumber\\
&+\Jad\Big[\v{T}_2\v{S}_1+\v{T}_1\v{S}_3+\tilde{\v{T}}_{1}\v{S}_{2L-2}+\tilde{\v{T}}_2\v{S}_{2L}\Big].
\label{ham_endspin_ad}
\end{align}
In the limit of large $\Jad$, it is obvious that the dimer configuration of  Fig. \ref{lattice_lrvbs}(a) is the ground state of Hamiltonian (\ref{ham_endspin_ad}). The DMRG result for the $\Jad$-dependence of the energy gap is shown in Fig. \ref{gapad}. This shows that the ground state for large $\Jad$  adiabatically continues to that with $\Jad=0$ without closing the energy gap. 

The distinction between spin-gap phases with different edge spins is ascribed to the topological difference in the bulk ground states  of Hamiltonian  (\ref{ham_open}).\cite{Pollmann2010,Pollmann2012,Zang2010,Chen2011,Hirano2008} According to ref. \citen{Chen2011}, only two topologically distinct spin-gap phases are possible in the spin-1/2 chains without translational invariance. 
\begin{figure}
\centerline{\includegraphics[width=7cm]{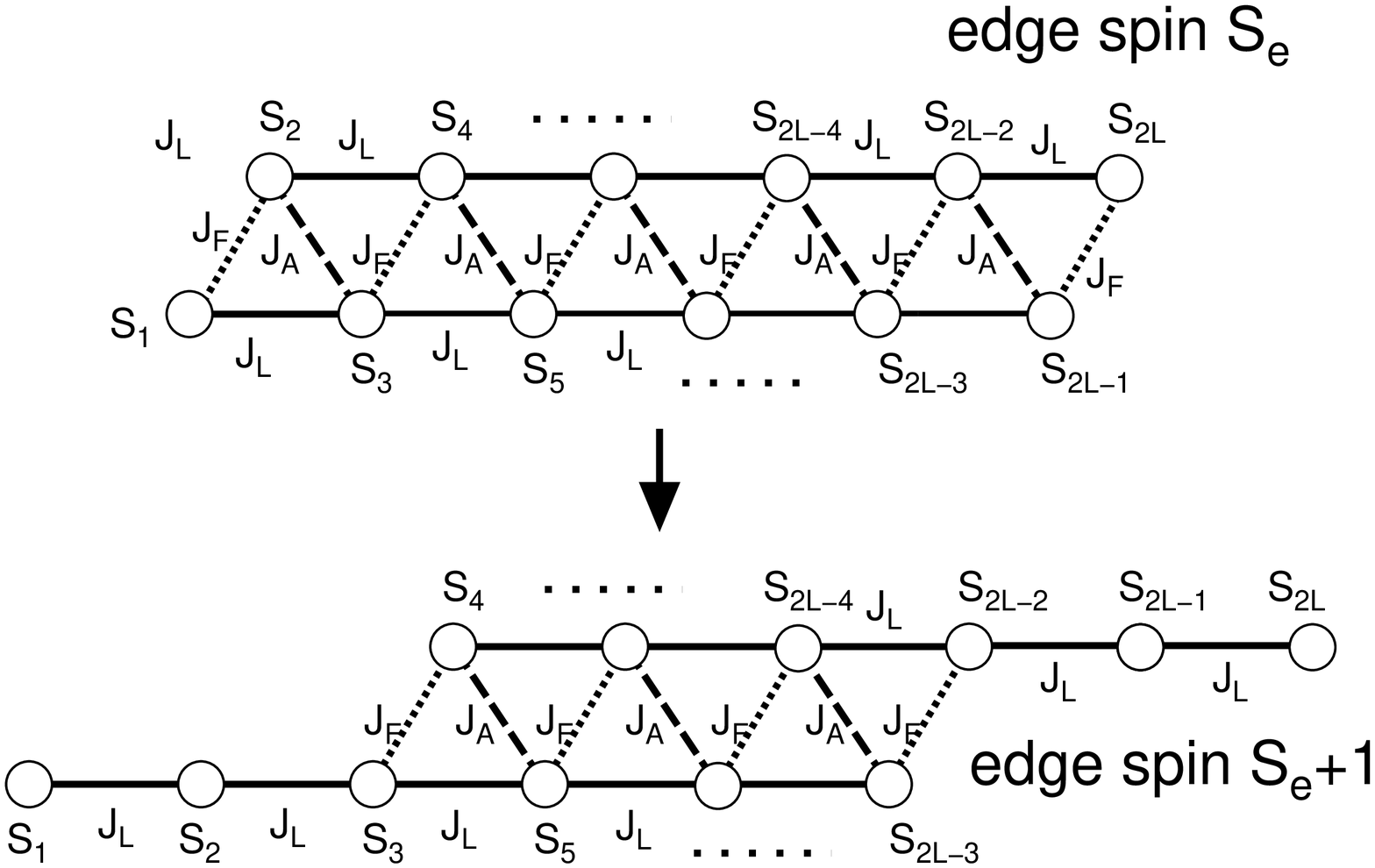}}
\caption{Local modification of $\HLo$.} 
\label{modify}
\end{figure}

 To see the above situation explicitly for the present model, we modify the Hamiltonian $\HLo$ 
into the Hamiltonian 
$\Hsu{mod}$ given by 
\begin{align}
{\Hsu{mod}} &=\sum_{l=2}^{L-1} \JF\v{S}_{2l-1}\v{S}_{2l}+\sum_{l=2}^{L-2} \JA\v{S}_{2l}\v{S}_{2l+1}\nonumber\\
&+\!\sum_{i=3}^{2L-4}\!\Jl\v{S}_{i}\v{S}_{i+2}+\Jl\v{S}_{1}\v{S}_{2}+\Jl\v{S}_{2}\v{S}_{3}\nonumber\\
&+\Jl\v{S}_{2L-2}\v{S}_{2L-1}+\Jl\v{S}_{2L-1}\v{S}_{2L},
\label{ham_mod}
\end{align}
 as shown in Fig. \ref{modify}. 
This modification consists of the  following local change of the exchange constants. The exchange constants of $\v{S}_{1}$-$\v{S}_{2}$ and ${\v{S}}_{2L-1}$-${\v{S}}_{2L}$ bonds are changed from $\JF$ to $\Jl$. Those of $\v{S}_{2}$-$\v{S}_{3}$ and ${\v{S}}_{2L-2}$-${\v{S}}_{2L-1}$ bonds are changed from $\JA$  to $\Jl$. Those of $\v{S}_{1}$-$\v{S}_{3}$, $\v{S}_{2}$-$\v{S}_{4}$, ${\v{S}}_{2L-2}$-${\v{S}}_{2L}$, and ${\v{S}}_{2L-3}$-${\v{S}}_{2L-1}$ bonds are set equal to 0. 
If the ground state of $\HLo$ 
has edge spin{s with magnitude} $\sed$, that of the modified Hamiltonian 
{$\Hsu{mod}$} has edge spin{s with magnitude} $\sed+1$ for ferromagnetic $\Jl$. {This implies that} 
 the $I_{\sed}$-phase and the $I_{\sed+1}$-phase {are actually the same phase}
, since this modification only concerns the local bonds and 
 the bulk ground state should remain unchanged, {because 
 it is unique and gapful, except for the quasi-degeneracy due to the edge spins,  as explained in \S\ref{interm}.
   In this context, all the {ground states} 
 with integer $\sed$ form a single phase, and those with half-odd-integer $\sed$ form another single phase. } {However, as discussed in \S\ref{strong}, it is arbitrary as to which of the two phases should be called ``topological'' and which one  ``trivial''. Therefore, we regard the I$_{\sed}$ phase with half-odd-integer $\sed$, connected to the Haldane phase of the spin-1 chain, the  ``topological'' phase and that with integer $\sed$ the ``trivial'' phase.} 
 Within the parameter space of the present model, however, we cannot move from the I$_{\sed}$ phase to the  I$_{\sed+1}$ phase  without passing through the I$_{\sed+1/2}$ phase.  Therefore, all I$_{\sed}$ phases with different $\sed$ are separate phases 
 in the present model.

We may speculate that  the I$_{\sed}$-I$_{\sed+1}$ phase boundary is a SU(2) symmetric Gaussian critical line with conformal charge unity, since this is the transition between two rotationally invariant spin-gap phases with different valence bond structures, such as the uniform point of the spin-1/2 isotropic dimerized Heisenberg chains. To confirm this speculation numerically, it is necessary to estimate the conformal charge from the ground-state energy with the periodic boundary condition obtained by numerical exact diagonalization. In the present case, however, this is difficult owing to the limitation of the tractable system size. 
 The intersections of the line of eq. (\ref{insta}) with the I$_{\sed}$-I$_{\sed+1}$ critical lines give the F-I$_{\sed}$-I$_{\sed+1}$ triple points. Obviously,  the line  of eq. (\ref{insta}) has no singularities at these triple points.

\subsection{Edge spin state on the ferromagnetic-nonmagnetic phase boundary}

\begin{figure}
\centerline{\includegraphics[width=6cm]{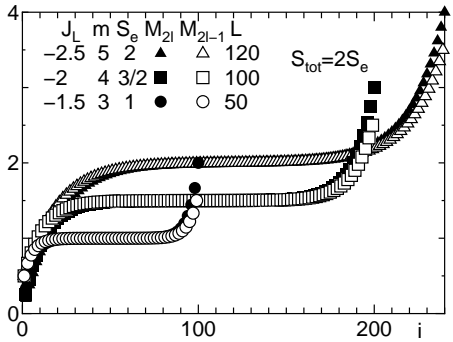}}
\caption{The  accumulated magnetization $M_i$ on the magnetic-nonmagnetic phase boundary (\ref{insta}) at the special points  $\Jl{/\JA}=-m/2 (m=3,4$, and 5) in the ground state with $\stot=\sztot=m-1$. The filled and open symbols represent the values for $i=2l$ and $2l-1$, respectively.}
\label{szav_sp}
\end{figure}
The exact ground state on the ferromagnetic-nonmagnetic phase boundary, eq. (\ref{insta}), is obtained by Dmitriev {\it et al.}\cite{dmitriev,dmitriev2} for the Hamiltonian
 \begin{align}
{\Hsu{ex}} &=\frac{\JF}{2}(\v{S}_{1}\v{S}_{2}+\v{S}_{2L-1}\v{S}_{2L})+\sum_{l=2}^{L-1} \JF\v{S}_{2l-1}\v{S}_{2l}\nonumber\\
&+\sum_{l=1}^{L-1} \JA\v{S}_{2l}\v{S}_{2l+1}+\sum_{i=1}^{2L-2}\Jl\v{S}_{i}\v{S}_{i+2},\label{ham_open_hf}
\end{align}
where the exchange couplings of the endmost bonds are halved to allow the exact solution.
The ground state for $\Jl/\JA\ne -m/2$, where $m$ is a positive integer, has a long-range spiral spin correlation with wavelength equal to the system size. This corresponds to the ferromagnetic state with a single twist. On the contrary, the ground state for $\Jl/\JA=-m/2$, which is called the ``special point'', has a short-range antiferromagnetic order. Therefore, it is expected that this solution continues to the I$_{\sed}$ states with the spin gap on the nonmagnetic side of the phase boundary. 

 To confirm this speculation, we examine the edge spins in the solutions at the special points. 
 Figure \ref{szav_sp} shows the  accumulated magnetization $M_i$ for total spin $\stot=\sztot=m-1$ calculated using the method in ref. \citen{Suzuki2008}. This figure shows that these solutions have edge spins with $\sed=(m-1)/2$ on  both ends 
 of the chain. As shown in the phase diagram in Fig. \ref{phase}, each special  point {with $m=2\sed+1$} belongs to the I$_{\sed}$ phase.  Thus, we may regard the special point  solution with $m$ as being representative of the I$_{\sed}$ phase for each $\sed=(m-1)/2$. Namely, our numerical calculation shows that there exists an I$_{\sed}$ phase of finite width around each special point. 

 Among the special point solutions, the solution with $m=2$ is an isolated dimer state on the $\JA$ bonds. This state remains the ground state all  along the line $\JF=2\Jl$ for $\Jl > -\JA$. This line belongs to the Haldane phase in the phase diagram of Fig. \ref{phase}. This confirms that the special point solution with $m=2$ is representative of the valence bond structure of the Haldane phase.

It appears pathological that the ground states for the special points $\Jl/\JA= -m/2$ and those for $\Jl/\JA\ne -m/2$ are totally different. However, it is known that the ground states are macroscopically degenerate in the thermodynamic limit along the magnetic-nonmagnetic phase boundary, eq. (\ref{insta}). This is analytically proven on the special points and  suggested by the numerical calculation for other points on this line\cite{dmitriev2}. From the continuity consideration, it is plausible that both types of eigenstates are actually present among the degenerate ground states for arbitrary values of $\Jl/\JA$ on the ferromagnetic-nonmagnetic phase boundary.

\section{Summary and Discussion}

Ground-state phases of the spin-1/2 ferromagnetic-antiferromagnetic alternating Heisenberg chain with ferromagnetic NNN interaction are investigated. 
In addition to the conventional Haldane phase and the ferromagnetic phase, we confirmed the presence of  intermediate spin-gap phases having edge spins with various magnitudes on both ends. These phases are classified into two topologically distinct phases according to whether the edge spin {$\sed$} is an integer or half-odd-integer {in the open-chain Hamiltonian (\ref{ham_open})}. {The physical picture of each phase is discussed on the basis of the appropriate modification of the Hamiltonian.} The relation to the exact solution on the ferromagnetic-nonmagnetic phase boundary is also discussed. It is found that each special point solution\cite{dmitriev,dmitriev2} with $\Jl/\JA=-m/2$ has edge spin $(m-1)/2$ and is representative of the I$_{\sed}$ phase with $\sed=(m-1)/2$.

Although we have characterized the nature of the spin-gap phases in our model as a series of topological phases by detecting the edge spins, several types of exotic phases are known to exist near the ferromagnetic-nonmagnetic phase boundary in 
other frustrated quantum spin chains. Hence, it is worthwhile to examine the possibility of their realization in the intermediate phases of the present model. 

One of the candidates is the spin nematic phase, which is a condensed state of bound multimagnons.\cite{Chubukov1991,lauchli,htsdec,khdlt}  In this case, the lowest bulk excitation should have total spin 2 or larger. We have checked numerically that this is not the case in the intermediate phases of the present model.

Spontaneously polymerized phases with broken translational symmetry are also candidates.\cite{hik.prb10,mg,ss,lauchli,Takano-K-S,hts} In the open chain, however,  one of the symmetry-broken states should be automatically selected if the system size is an integer multiple of the periodicity of the ground state. Hence, a spatial modulation of physical quantities should be observed in the ground state calculated by DMRG. Such a spatial modulation is not observed in the intermediate phases of the present model. 

The spiral (quasi-)long-range ordered phase is also excluded because of the presence of the bulk spin gap. 
Nevertheless, the  short-range spiral order  might be possible even in  the spin-gap phases. In this case, the spin gap should open at a finite wave number.\cite{nomura_ic} Although we found no evidence of such behavior within the system size accessible by the numerical exact diagonalization method, the possibility of a spiral short-range order with pitch longer than the system size remains possible. Even in this case, the ground-state phase transition does not take place as long as the spiral order remains short ranged.

The physical origin of the edge spin with general values of $\sed$ may be understood in the following way. For large ferromagnetic $\Jl$, the spins on the odd-th (even-th) sites are strongly correlated ferromagnetically with each other. In the nonmagnetic phase, however, this correlation does not extend over the whole chain, but is cut into finite ferromagnetic clusters fluctuating in position and length. The clusters on the even-th sites and odd-th sites are correlated antiferromagnetically,  forming a nonmagnetic ground state as a whole. On both ends of the chain, the endmost clusters are pinned to the edges, resulting in the edge states with spin $\sed$. 

The ground-state phase diagram is successfully determined within the present analysis, which is mainly based  on the energy gap. Recently, however, it was proposed to characterize the topological difference in the spin gap phases by the quantized Berry phase\cite{Hirano2008}, 
the degeneracy of the entanglement spectrum, and the space inversion parity\cite{Pollmann2010,Pollmann2012}. It is possible that the numerical and analytical investigations of these quantities  can lead to a deeper understanding of the nature of the I$_{\sed}$ phases and  the corresponding special points. These tasks are left for future studies.

The numerical diagonalization program is based on the package TITPACK ver.2 coded by H. Nishimori.  The numerical computation in this work has been carried out using the facilities of the Supercomputer Center, Institute for Solid State Physics, University of Tokyo, the Supercomputing Division, Information Technology Center, University of Tokyo, and   Yukawa Institute Computer Facility, Kyoto University.  This work is supported by a Grant-in-Aid for Scientific Research (C) (21540379) from the Japan Society for the Promotion of Science.

\end{document}